# Numerical Study of Structural Phase Transitions in a Vertically Confined Plasma Crystal

K. Qiao and T. W. Hyde
*Center for Astrophysics, Space Physics and Engineering Research*
*Baylor University, Waco, TX, 76798-7310, USA*
Ke_Qiao@Baylor.edu    Truell_Hyde@Baylor.edu

**Abstract**

Dusty plasmas consists of an ionized gas containing small (usually negatively charged) particles. Dusty plasmas are of interest in both astrophysics and space physics as well as in research in plasma processing and nanofabrication. In this work, the formation of plasma crystals confined in an external one-dimensional parabolic potential well is simulated for a normal experimental environment employing a computer code called BOX_TREE. Such crystals are layered systems, with each layer a two dimensional lattice composed of grain particles. The number of layers is dependent upon the external potential parameter. For constant layer number, the intralayer structure transits from a square lattice to a hexagonal (triangular) lattice as the confining potential decreases. For hexagonal lattices, both hcp and fcc characteristics were found but hcp structures dominate. The relative thickness of the system was also examined. The results were compared with previous experimental and theoretical results and found to agree.

## 1. INTRODUCTION

Dusty plasmas play an important role in astrophysical environments. Although there is little or no evidence yet of strongly coupled dusty plasmas in nature, the formation and stability mechanisms for ordered colloidal crystals within a tenuous dusty plasma is of great interest in protoplanetary, protostellar and accretion disk formation as well as spiral galaxies and dark matter research. A plasma crystal under typical experimental environments is formed when dust particles are levitated in an rf discharge plasma sheath above the lower electrode (Chu and I, 1994; Hayashi and Tachibana, 1994; Thomas et al., 1994). The particles are negatively charged with this charge shielded by the ambient plasma; thus the interaction between particles is a repulsive Yukawa potential defined by

$$v(r) = q \exp(-r/\lambda_D)/4\pi\varepsilon_0 r, \qquad (1)$$

where $q$ is the particle charge, $r$ is the distance between any two particles and $\lambda_D$ is the dust Debye length.

The total external potential in the plasma sheath has been shown to be a parabolic potential well (Tomme et al., 2000), thus it can be modeled as (Totsuji, 1997)

$$v(z) = \frac{\mu}{2}z^2 \tag{2}$$

where $z$ is the particle height and $\mu$ is the parabolic coefficient.

Structural phase transitions have been investigated by Dubin (1993) for a vertically confined One Component Plasma (OCP) system and by H. Totsuji (1997) for plasma crystals employing a confined Yukawa system as a model. These structural phase transitions have been observed experimentally in OCP systems (Mitchell et al., 1998) and colloidal suspensions (Van-Winkle and Murry, 1986). A phase diagram has been established for the confined Yukawa system characterized by two dimensionless parameters, $\kappa$ and $\eta$ (Totsuji, 1997). The shielding parameter $\kappa$ is defined by

$$\kappa = \frac{a}{\lambda_D}, \tag{3}$$

where $a$ is the mean distance between particles defined by $N_s = 1/\pi a^2$ with $N_s$ the surface number density in the $xy$ plane. $\eta$ is defined by

$$\eta = \frac{\mu}{4\pi q^2 N_s^{3/2}}. \tag{4}$$

In this research, a numerical code called Box_Tree (Richardson, 1993; Vasut and Hyde, 2001; Matthews and Hyde, 2003; Qiao and Hyde, 2003) is used to simulate the crystallization of a vertically confined complex plasma modeled as a Yukawa system. The plasma crystal structure is examined in detail and the structural phase transitions, including the transition between crystals with different numbers of layers and with the same number of layers but different intralayer structures, is investigated and compared with previous theoretical and experimental research results.

## 2. SIMULATION METHODS

The dust particles in the system considered have a constant and equal charge $q = 3.84 \times 10^{-15} C$, equal mass $m_d = 1.74 \times 10^{-12} kg$ and a radius $r_0 = 6.5 \mu m$. The interparticle interaction is assumed to be a Yukawa potential with a Debye length $\lambda_D = 0.75 mm$ and the external potential is modeled as a parabolic potential given by Eq. (2). The box size is set at $15 \times 15 \times 15 mm^3$ and 600 particles are considered; thus the surface number density $N_s$ the mean distance $a$, and the shielding parameter $\kappa$ are equal to $2.67/mm^3$, $0.346mm$ and 0.61, respectively. The neutral gas drag is included with the Epstein drag coefficient (Epstein, 1923) set at $\beta = 2.22 s^{-1}$. Periodic boundary conditions in the XY direction are employed since the system considered has a size much smaller than that of a plasma crystal under a typical experimental environment. On the Z direction, a closed boundary condition is used with particles impacting the top or bottom boundaries of the box reflected under an elastic collision.

All simulations start with a random distribution of particles placed in the box subject to the condition that the system's center of mass must be located at the center of the box. The initial velocities of the particles are all set to zero. An ordered lattice (plasma crystal) forms approximately 65s after the start of the simulation.

## 3. RESULTS

Simulations were conducted for 46 η values between 0.0034 and 0.48. Over this range, the system transits from a single-layer crystal to a two, three, four and five-layer crystal as shown in Fig.1. These transitions start with the system forming a single-layer crystal for the range $0.48 \geq \eta \geq 0.456$. As η decreases, the system evolves through a 1-2 layer transition at approximately 0.336. The system remains in a two-layer state for $0.336 \geq \eta \geq 0.072$, a three-layer state for $0.06 \geq \eta \geq 0.0216$, a four-layer state for $0.0204 \geq \eta \geq 0.0084$ and a five-layer state for $0.0072 \geq \eta \geq 0.0034$. This is to be expected since as can be seen in Eq. (4), a decrease in η causes a decrease in the confining potential.

As shown in Figures 2 and 3, inside each stage as identified above, although the number of layers remains constant, the intralayer structure changes from that of a square lattice to that of a hexagonal (triangular) lattice as η decreases (or as the confining potential decreases). It can also be seen in Figs. 2 and 3 (b) and (e) that between each of these structure types, a transition stage exists where the lattice exhibits a complex structure showing a mixture of both square and hexagonal symmetry. This agrees with both theoretical predictions (Dubin, 1993; Totsuji, 1997) and experimental results (Mitchell et al., 1998; Van-Winkle and Murry, 1986).

Figures 4 and 5 show the relationship between actual particle positions in different layers for crystals with differing number of layers. For both square and hexagonal structures, the layers stack in a staggered fashion, with the upper particles immediately above the center of the corresponding lattice cell below. As shown in Figures 4 and 5, b and d, particles on the third layer are primarily aligned vertically with particles in the first layer for hexagonal lattices, which is a characteristic of a hcp lattice. For a hcp lattice, particle positions within a hexagonal lattice plane repeat themselves every other plane; thus the planes are ordered as ABABAB …. On the other hand, for a fcc lattice, particle positions repeat every three planes; thus the planes are ordered as ABCABC …. As seen in Figures 4 and 5, the majority of lattices in this simulation are hcp even though only a portion of the lattice for the four-layer crystal shows a fcc characteristic. This is in agreement with both the OCP (Mitchell et al., 1998) and the colloidal suspension (Van-Winkle and Murry, 1986) experiments. The vertical alignment of particles in consecutive layers observed in plasma crystal experiments on earth (Thomas et al., 1994) was not seen in these simulation results. However, this is to be expected since such vertical alignment is caused by the wake effect of the ion flow and the system considered in this research is a pure Yukawa system.

The relative thickness of the system, calculated as the absolute distance between the top and bottom layer and then normalized by the mean distance *a*, was also investigated as a function of η and is shown in Fig.6. It can clearly be seen that the relative thickness of the system increases as η decreases and that there are discontinuities in the function corresponding to the stepwise transitions in the number of layers N. The dependence of the intralayer structures on η is also shown in Fig.6. Both the structural phase transitions and the d-η function agree with Totsuji's predictions (Totsuji, 1997) quantitatively.

## 4. CONCLUSIONS

In summary, the crystallization of a complex plasma modeled as a vertically confined Yukawa system was simulated using the Box_Tree code. The system was found to exist as layered crystals with a different number of layers for different confining potentials. When the number of layers remains constant, the intralayer structure transitions from that of a square lattice to that of a hexagonal (triangular) lattice as the confining potential becomes weaker. This is in agreement with previous theoretical and experimental results. For hexagonal lattices, both hcp and fcc symmetry was seen with hcp structure dominating. This agrees with both OCP and colloidal suspension experiments while the d-η function shown was found to agree with Totsuji's predictions quantitatively.

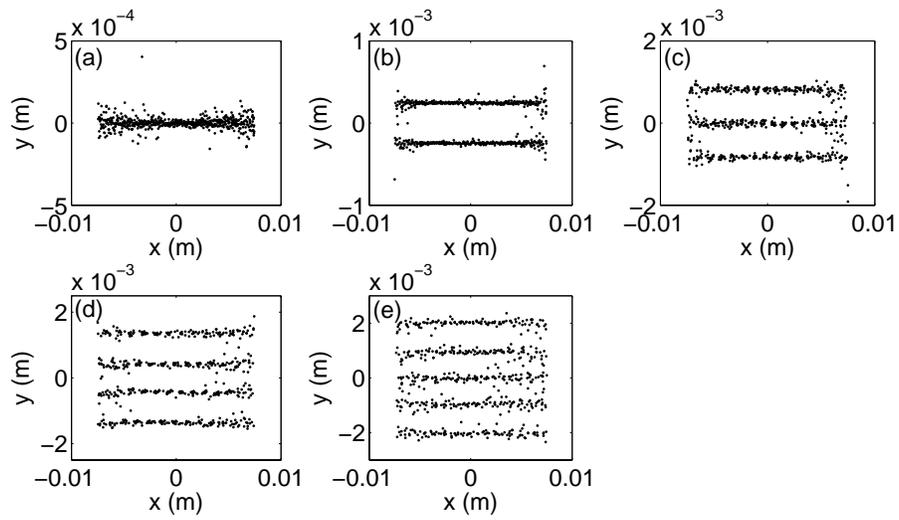

Fig. 1. Side view of the (a) single-layer, (b) two-layer, (c) three-layer, (d) four-layer and (e) five-layer crystal as η decreases..

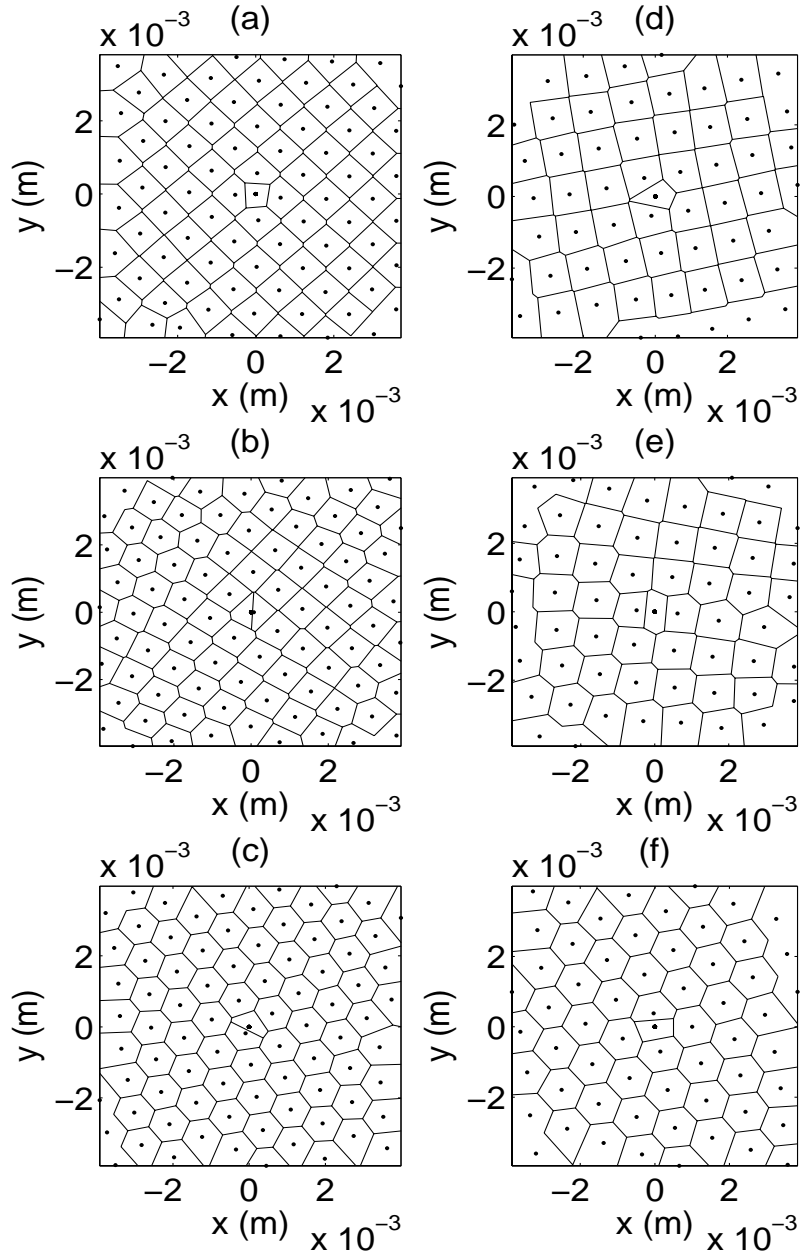

Fig. 2. Voronoi diagram for the lattice in the middle layer (one of the middle layers for an even numbered system) of a vertically confined Yukawa system when (a) $\eta = 0.336$, (b) $\eta = 0.168$, (c) $\eta = 0.096$ (two-layer system), (d) $\eta = 0.06$, (e) $\eta = 0.042$ and (f) $\eta = 0.0216$ (three-layer system).

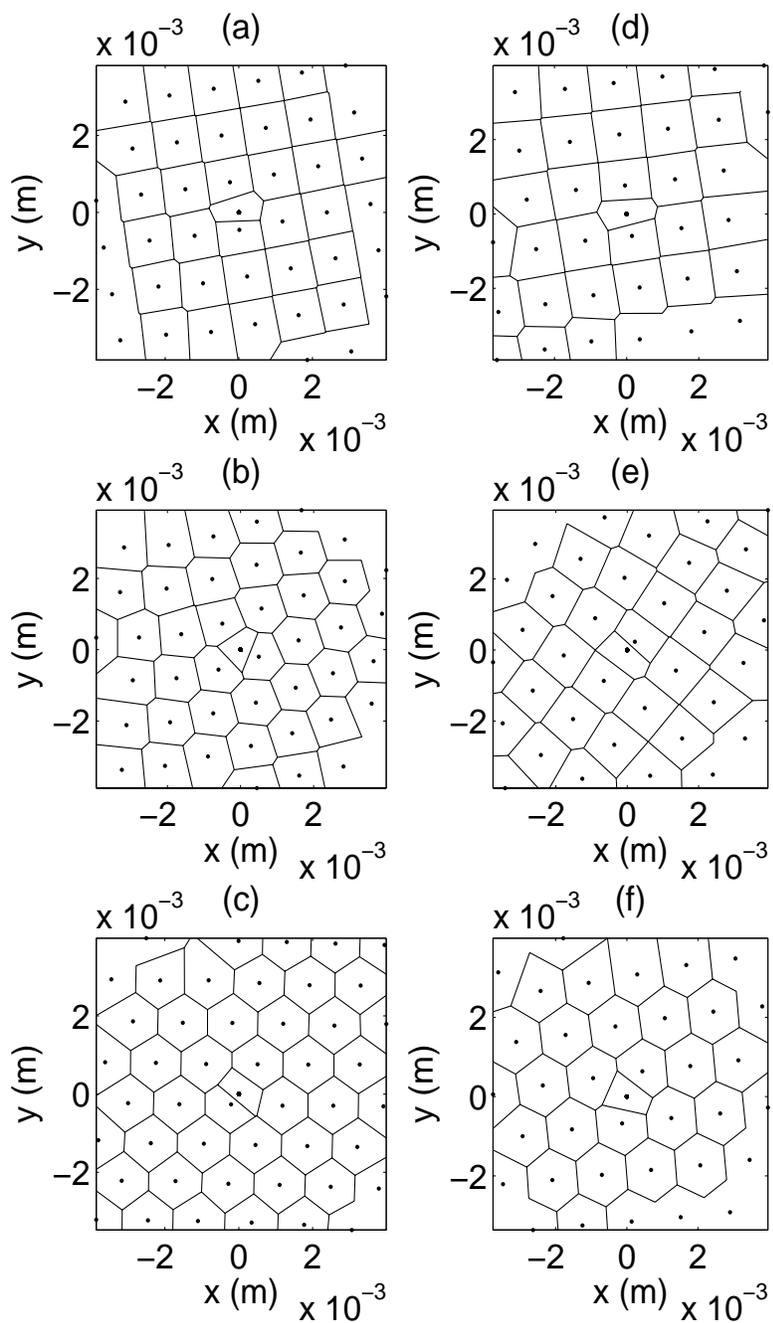

Fig. 3. Same as Fig. 2 but with (a) $\eta = 0.0204$, (b) $\eta = 0.0132$, (c) $\eta = 0.0084$ (four-layer system), (d) $\eta = 0.0066$, (e) $\eta = 0.0054$ and (f) $\eta = 0.0034$ (five-layer system).

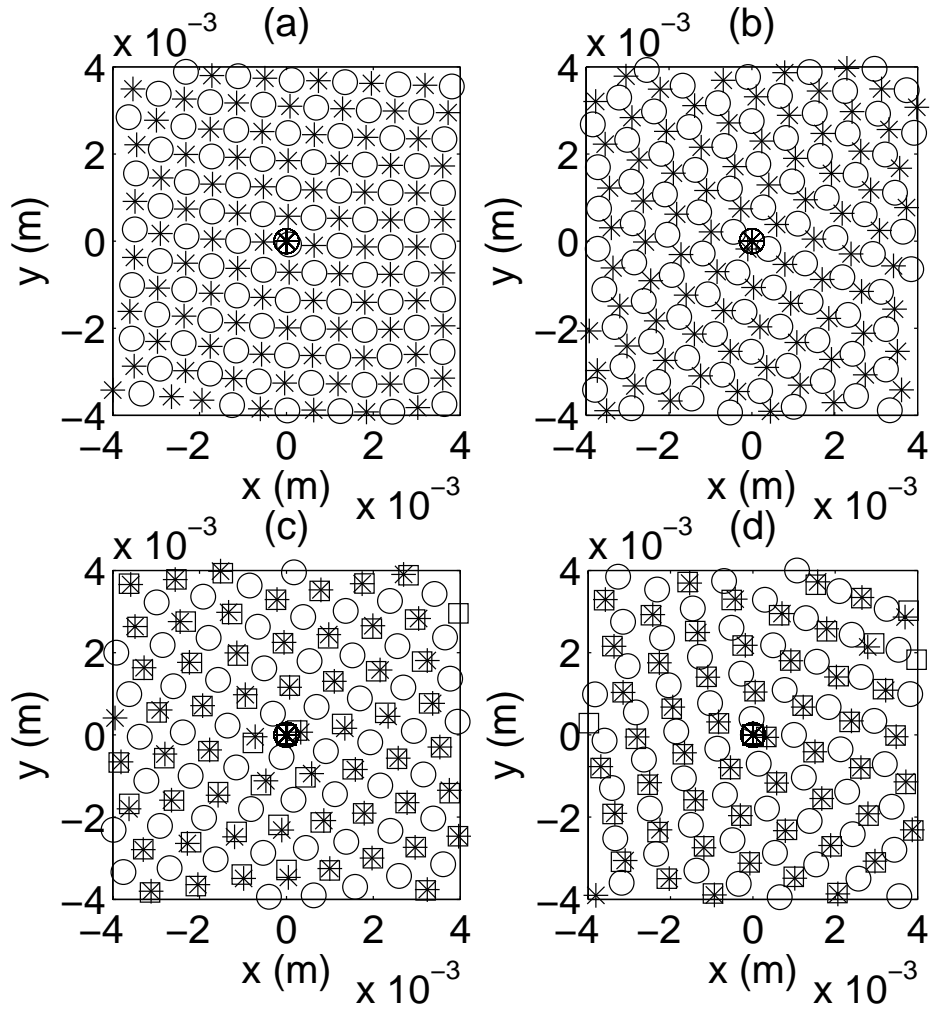

Fig. 4. Top view of a vertically confined Yukawa system when (a) $\eta = 0.336$, (b) $\eta = 0.096$ (two-layer system), (c) $\eta = 0.06$ and (d) $\eta = 0.0216$ (three-layer system). The asterisks, circles and triangles represent particles in the first, second and third layers, respectively.

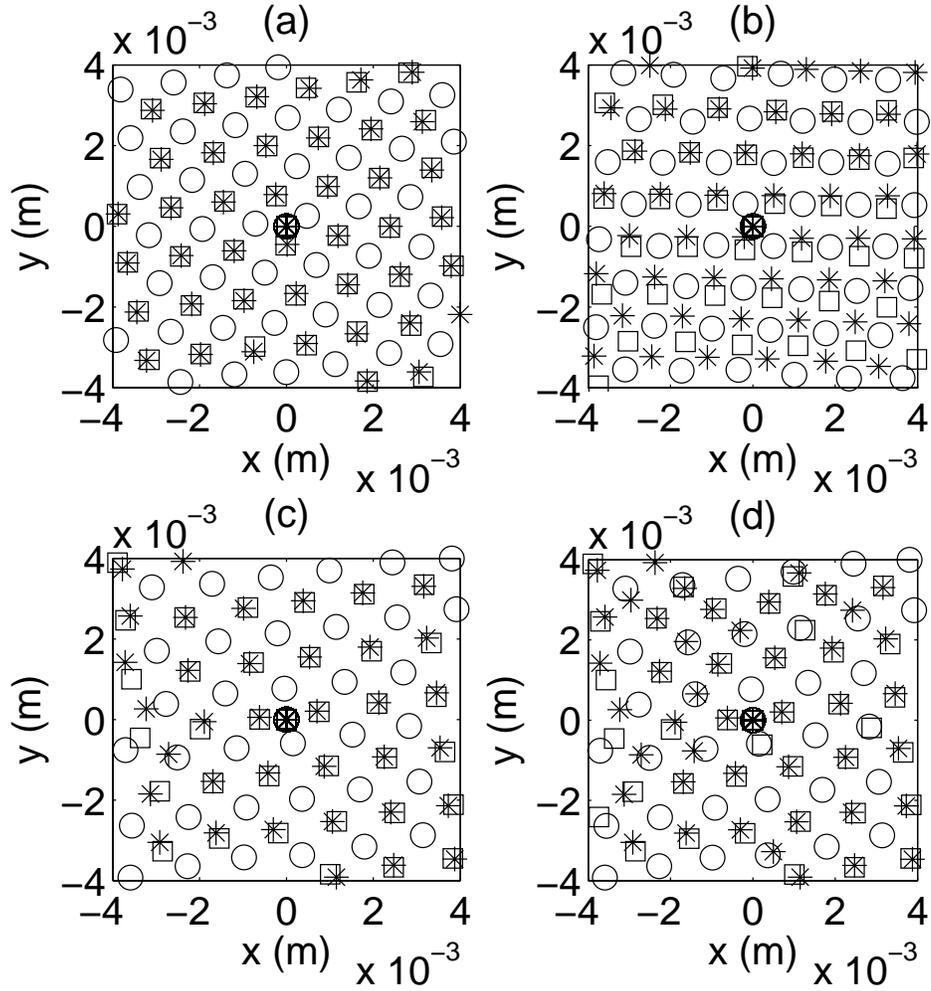

Fig. 5. Same as Fig. 4 but with (a) $\eta = 0.0204$, (b) $\eta = 0.0084$ (four-layer system), (c) $\eta = 0.0066$ and (d) $\eta = 0.0034$ (five-layer system).

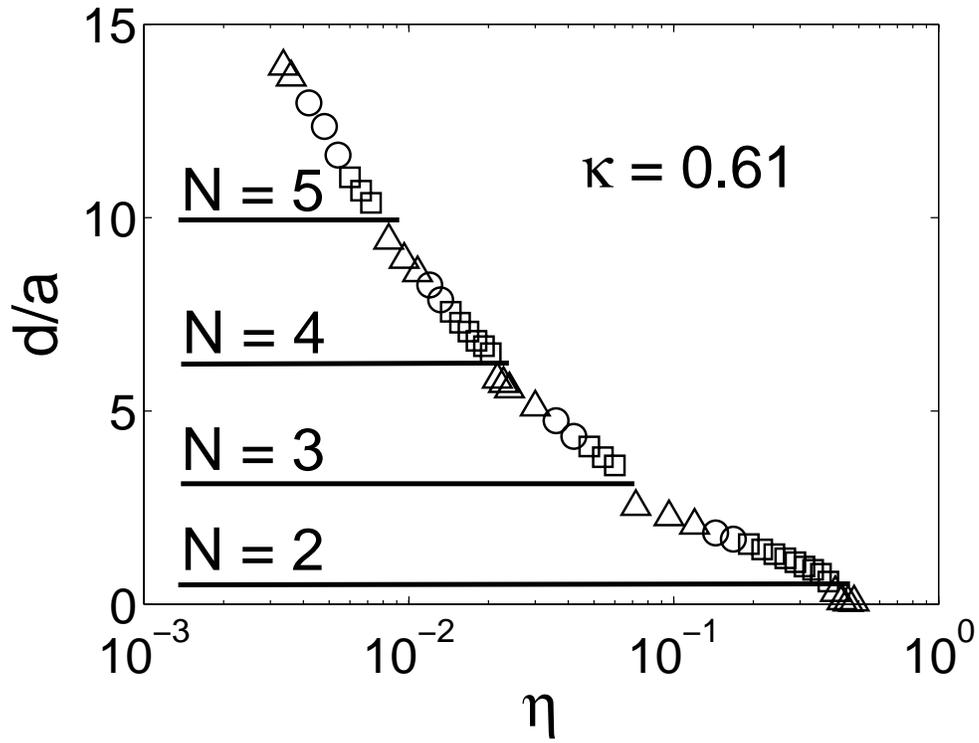

Fig. 6. The system's relative thickness $d/a$ as a function of the characteristic parameter $\eta$. The intralayer structure with square, triangular and complex symmetries are represented by squares, triangles and circles respectively.